\documentclass[acmsmall,screen,review=false,anonymous=false,natbib=false]{acmart}
\setcopyright{cc}
\copyrightyear{2025}
\acmYear{2025}
\acmDOI{}

\usepackage[T1]{fontenc}
\usepackage[utf8]{inputenc}
\usepackage{textcomp}
\usepackage{newtxtext}
\usepackage{newtxmath}

\usepackage{microtype}
\usepackage{alltt}
\usepackage{fancyvrb}
\usepackage{booktabs}
\usepackage{stmaryrd}
\usepackage{graphicx}
\usepackage[capitalize,noabbrev,nameinlink]{cleveref} 
\usepackage{csquotes} 
\usepackage{amsmath}
\usepackage{caption}
\usepackage{listings}
\usepackage{todonotes}
\usepackage{tcolorbox}
\usepackage{simplebnf}
\usepackage{subcaption}
\usepackage{tikz}
\PassOptionsToPackage{unicode=false}{hyperref}

\usepackage[
  backend=biber,
  natbib=true,
  sortcites=true,
  maxcitenames=2, 
  maxbibnames=4,  
  minbibnames=3,  
  datamodel=acmdatamodel,
  style=acmnumeric,
  language=american
]{biblatex}
\lstdefinelanguage{mk}{
  morekeywords={==, fresh, conde, defrel, run, run*},
  sensitive=true,
  morecomment=[l]{;},
}
\DefineVerbatimEnvironment%
  {VerbatimEsc}{Verbatim}
  {commandchars=\\\{\}}

\usepackage{mathpartir}

\newcommand{\prg}[2]{\mathtt{prg}\left(#1\right)\;#2}
\newcommand{\step}[1]{\ensuremath{\Rightarrow^{\text{\textsc{#1}}}}}
\newcommand{\delay}{\ensuremath{\mathtt{delay}}}
\newcommand{\go}{\ensuremath{\mathtt{go}}}

\renewcommand{\Top}{\ensuremath{\top}}
\newcommand{\answer}{\ensuremath{(\Top\,\sigma)}}
\newcommand{\strcat}{\ensuremath{\mathtt{'cat}}}
\newcommand{\strdog}{\ensuremath{\mathtt{'dog}}}
\begin{document}

\title{Visualizing miniKanren Search with a Fine-Grained Small-Step Semantics}

\begin{abstract}
  We present a deterministic small-step operational semantics for miniKanren that explicitly represents the evolving search tree during execution.
  This semantics models interleaving and goal scheduling at fine granularity, allowing each evaluation step—goal activation, suspension, resumption, and success---to be visualized precisely. Building on this model, we implement an interactive visualizer that renders the search tree as it develops and lets users step through execution.
  The tool acts as a pedagogical notional machine for reasoning about miniKanren’s fair search behavior, helping users understand surprising answer orders and operational effects.
  Our semantics and tool are validated through property-based testing and illustrated with several examples.
\end{abstract}

\author{Brysen Pfingsten}
\email{pfingsbr@shu.edu}
\orcid{0009-0000-7634-1045}
\affiliation{%
  \institution{Seton Hall University}
  \country{USA}
}

\author{Jason Hemann}
\email{jason.hemann@shu.edu}
\orcid{0000-0002-5405-2936}
\affiliation{%
 \institution{Seton Hall University}
 \country{USA}
}

\maketitle


\section{Introduction}\label{sec:introduction}

The miniKanren family of logic programming languages uses a fair interleaving depth-first search strategy~\cite{kiselyov2005backtracking} to find solutions.
The execution of a logic program involves a complex, non-deterministic search process, making it difficult for a programmer to predict the order in which a given program will search or how performant that search will be.
Small differences in goal ordering or structuring can dramatically alter the order in which answers appear or the time it takes to compute them.
Implementers too may find it difficult to anticipate the consequences of even slightly altering where their implementation places delays.
For example, consider the miniKanren query in \cref{fig:concrete-and-abstract}.

\begin{figure}[htbp]
    \small
  \begin{minipage}[t]{0.45\textwidth}
\begin{verbatim}
(defrel (same x y) (== x y))

(run* (q)
  (conde
    [(conde
       [(same q 'turtle)]
       [(same q 'cat)]
       [(== q 'dog)])]
    [(same q 'fish)]))
\end{verbatim}
  \end{minipage}
  \begin{minipage}[t]{0.44\textwidth}
\begin{verbatim}
same(X,Y) :- X = Y.

?- ( ( same(Q,turtle)
     ; same(Q,cat)
     ; Q = dog
     )
   ; same(Q,fish)
   )
\end{verbatim}
  \end{minipage}
  \caption{Definitions and uses of the \texttt{same} relation, written in both \citet{reasoned2}'s concrete miniKanren syntax as well as in Prolog.}\label{fig:concrete-and-abstract}
\end{figure}

Intuitively, a programmer might expect the answer \texttt{turtle} to appear first, because it appears earliest in the source code.
The Prolog implementation produces \texttt{Q = turtle}, \texttt{Q = cat}, \texttt{Q = dog}, and \texttt{Q = fish} in that order.
Running this query in a miniKanren like those described by \citet{reasoned2} or \citet{nearlymacrofree}, however, produces \texttt{'(fish turtle dog cat)}.
In practice, miniKanren can easily produce answers in an unintuitive order due to its fair scheduler and interleaving strategy.

Without a clear understanding of the operational semantics, the reasons behind such behavior remain obscure.
Declarative debugging techniques~\cite{declarativedebugging,caballero2017survey} offer limited help here, as the issue is not logical correctness---all branches eventually succeed---but rather the operational details of how the search is conducted.

In this paper, we address precisely this challenge by introducing a \textbf{deterministic small-step operational semantics} for miniKanren that \emph{explicitly encodes the evolving search tree}.
We model the nondeterministic miniKanren computation as deterministic transformations of a search tree.
Unlike existing formal semantics for logic programming languages~\cite{kowalski1971linear,de1990comparative}, our approach models miniKanren's fair interleaving, and does so at a \emph{fine-grained, intra-goal} level.
More specifically, our semantics exposes each distinct evaluation step within the execution of individual miniKanren goals, including goal initiation, suspension, resumption, and final success or failure.
Previous work on miniKanren semantics either models interleaving at a higher, coarser granularity---stepping directly from one completed relational call to another without exposing the crucial internal decision points---or uses a simpler approximation of typical implementations' search behavior~\cite{rosenblatt2019firstorder,rozplokhas2019certified}.
Our approach, by contrast, transforms the search tree at each intra-goal evaluation step, (clarifying when, how, and why each goal is executed, paused, or resumed), and models the more parsimonious interleave behavior of existing implementations.

We use these semantics as the basis for an interactive \textbf{search-tree expression stepper}, inspired by expression steppers from the programming languages education literature~\cite{lieberman1984steps,clements2001modeling,culpepper2007debugging,clements2022towards}.
This tool functions as a pedagogical \emph{notional machine}~\cite{boulay1981black} for miniKanren, explicitly visualizing the evolving search tree at every evaluation step.
Students and new users can interactively step forward and backward through the search process, inspecting the current state, suspended goals, and chosen branches.
We hope that in doing so, they build an accurate mental model of how miniKanren conducts its fair interleaving search and better understand how logic programs execute.
While primarily designed as an educational visualizer, our tool may also provide experienced miniKanren programmers with limited diagnostic capabilities.
The prototype is available online (\href{http://minikanrenredex-prod.shu.edu}{minikanrenredex-prod.shu.edu}) for the community to test and use.

Using PLT Redex~\cite{pltredex,klein2012run}, we built a \textbf{mechanized implementation} of our semantics.
Our tool is built around this mechanization, which we use as an executable reference interpreter.
This mechanization is also a platform for automated testing of key properties like determinism and preservation of well-formedness.

To demonstrate our semantics and visualization approach concretely, we present detailed stepping traces of miniKanren queries, including the unintuitive ordering scenario above.
Our semantics framework can accommodate alternate search strategies; we validate this by also implementing a Prolog-style depth-first search by simply swapping out the reduction rules.
We believe this approach provides a valuable pedagogical foundation for teaching miniKanren's operational behavior, as well as a sound semantic basis for future exploration of logic programming execution strategies.

The remainder of the paper proceeds as follows.
\Cref{sec:background} reviews miniKanren's syntax, gives a high-level intuitive comparison of miniKanren's operational semantics to that of Prolog, and reprises interleaving search strategies.
\Cref{sec:language-semantics} defines our formal small-step semantics explicitly, explaining the search-tree representation and detailed reduction rules.
\Cref{sec:mixed-redex-js} describes our interactive visualizer tool's design goals, implementation, its integration with the Redex semantics, and our design choices to improve usability.
In \Cref{sec:visu-reduct}, we illustrate the semantics and visualizer in action with concrete miniKanren queries, showing how stepping through execution can illuminate otherwise confusing search behaviors.
Finally, \Cref{sec:related-work} situates our contributions within existing work on logic programming visualization and semantics, and \Cref{sec:conclusion} discusses future directions and potential extensions of this approach.

\section{Background}\label{sec:background}\label{sec:our-minik-lang}

In this \lcnamecref{sec:background}, we discuss some background and ideas from CS education research.
We revisit traditional Prolog traces and the Byrd box model.
We also re-introduce miniKanren and its semantics informally, by way of a contrast with Prolog and traditional logic programming.

\subsection{Visual steppers and notional machines}

Students' difficulty in understanding logic program execution---what is sometimes referred to as the \enquote{Prolog story} problem---has been well documented in computing education research~\cite{taylor1987studying,pain1987what,stekovanic2022challenges}.
Learners often struggle to form mental models of a language's non-deterministic control flow, especially backtracking, goal scheduling, and the order of answer production.
Over the years, researchers and educators have explored a wide variety of tracers and visualization tools to help make the execution process more accessible.

A \emph{notional machine}~\cite{boulay1981black} is a simplified conceptual model used in education to help learners reason about program execution.
Similarly, a \emph{visual stepper} is a pedagogical tool allowing users to step through program evaluation visually, forming correct mental models by observing the evolving state.
Tools like the PLT Redex stepper have proven valuable pedagogically precisely because they expose each small step of computation explicitly and clearly.

\subsection{Traditional Prolog execution models and debugging}

Early Prolog debuggers were built around the box model, introduced by \citet{byrd1980understanding} as a way to help users understand Prolog's control flow.
In the four-port Byrd model, each predicate call is conceptualized as a \enquote{box} that goes through ports: typically Call, Exit (success), Fail, and Redo.
This model underlies most Prolog tracers allowing the programmer to single-step through each call and backtracking event.
The Byrd box model's simplicity made it ubiquitous, but it can produce very verbose traces and requires the user to mentally reconstruct the program's logical structure from low-level steps.
To provide higher-level views of execution, researchers turned to AND/OR tree representations.
An AND/OR tree depicts the search space of a query as a tree of goals: an AND-node's children (placed side-by-side) are all the sub-goals of a conjunction, and an OR-node's children are alternative goal expansions.
Each path down an AND/OR tree corresponds to a potential solution, and backtracking corresponds to exploring alternate branches (OR-nodes) of the tree.
Tree models like this make backtracking points and the overall search structure explicit, which is harder to see in the linear Byrd port trace.
However, they can become large and cluttered for complex programs, and relating them back to code (and features like the cut) is non-trivial.

\subsection{miniKanren's simplified execution model}

The miniKanren language is simpler---both more verbose and less expressive---than more traditional logic or constraint-logic programming languages.
It differs from traditional Prolog-style logic languages in several key ways that together simplify the language's execution model, making it particularly amenable to a fine-grained, explicit semantics as we develop in \cref{sec:language-semantics}.
The language assumes a fixed set of relational definitions with no dynamic \texttt{assert} or \texttt{retract}.
Calls to undefined relations or relation-arity mismatches are treated as errors, and can in some cases be caught statically by the host language.
Unlike Prolog, miniKanren has no implicit head unification for relations.
The programmer writes each predicate directly in a form of program completion~\cite{clark1978negation,lloyd1984making}.
Relations are invoked via explicit parameter substitution (akin to $\beta$-substitution in functional languages), and all unification is done explicitly via goals.

The miniKanren term language is first-order, meaning relations themselves are not in the domain of terms and there are no higher-order relations.
The programmer explicitly introduces new logic variables.
Lexical variables (local to a relation) and logical variables (introduced implicitly by \texttt{fresh}) have clearly disjoint scopes.
Associating a lexical variable with its corresponding logic variable can also be modeled by $\beta$-substitution.

The core language is small enough that the example in \cref{fig:concrete-and-abstract} exercises most of the miniKanren language forms.
Negation is generally disallowed, and in most dialects limited to a handful of negated primitive constraints on terms~\cite{lassez89independence,hemann2020constraint}.
The miniKanren programming community emphasizes pure, cut-free declarative programming, which aligns with many modern Prolog best practices~\cite{triska2025power}.
Even the miniKanren arithmetic suite~\cite{kiselyov2008pure} is implemented as library routines over bit-vectors that behave reasonably in all modes.
Host-language macro extension facilities are used to provide less cumbersome surface forms that de-sugar to the core.
User-level host-language macros scratch much the same itch as meta-programming in Prolog.

\section{Language and Semantics}\label{sec:language-semantics}

In this section we present our model of miniKanren execution, which we define by a reduction semantics.
We assume familiarity with standard reduction semantics terminology and notation, as described in, e.g., the PLT Redex textbook~\cite{pltredex}.
Readers primarily interested in the visualization aspects or high-level intuition may skip ahead without significant difficulty.

Unlike typical miniKanren implementations that use implicit streams of answers, our semantics encodes the entire search tree and its relation environment in a single expression.
The main technical contribution described in this section is treating a nondeterministic search process as a deterministic state transition system over a tree.
We get a deterministic single-step transition that still models nondeterministic search, by threading the whole search tree through states.
We first describe the grammar of programs, relations, and search trees, then explain the reduction rules and how they realize the standard miniKanren search strategy.

\subsection{Language}\label{sec:language}\label{sec:eval-contexts}

\Cref{fig:language} shows the grammar of our core language.
Our grammar consists of a core microKanren-like language extended with explicit constructs to represent the state of the search: notably, a search tree (S) data structure is a component of
the program.
A program is a triple \texttt{prg} $\Sigma$ $S$ where $\Sigma$ is the \textit{relation environment} and $S$ is a \textit{search tree}.
As usual each relation definition in $\Sigma$ is given as $r(L) \leftrightarrow G$ with $r$ a relation name, $L$ a list of syntactic variables (parameters), and $G$ a goal that constitutes the relation's body.
\begin{figure}[htbp]
    \small
  \begin{minipage}[t]{0.45\textwidth}
    \begin{align*}
      \textsc{Programs} \quad & \ni p \coloneqq \texttt{prg}\;\Sigma\; S\\
      \textsc{Environments} \quad & \ni \Sigma \coloneqq r(L) \leftrightarrow G \dots\\
      \textsc{Parameter Lists} \quad & \ni L \coloneqq x \dots\\
      \textsc{Search Trees} \quad & \ni S \coloneqq \texttt{empty} \mid G\;\sigma \mid S \rightarrow S \mid S \leftarrow S\\
        &\qquad\mid S + S \mid S \times G\\
        &\qquad\mid \texttt{go}\; S\\
        &\qquad\mid \texttt{delay}\; S\\
      \textsc{Goals} \quad & \ni G \coloneqq \top \mid t \equiv t \mid r(t\dots) \\
        &\qquad \mid G \vee G \mid G \wedge G \mid \exists\;(L)\;G\\
    \end{align*}
  \end{minipage}
  \begin{minipage}[t]{0.44\textwidth}
    \begin{align*}
      \mathbb{N} \quad &\ni i\\
      \textsc{Constants} \quad &\ni c\\
      \textsc{Variables} \quad &\ni x, y\\
      \textsc{Relation Names} \quad &\ni r\\
      \textsc{Logic Variables} \quad &\ni l_i\\
      \textsc{Terms} \quad &t \coloneqq c \mid x \mid l_i \mid t\colon t\\
      \textsc{States} \quad &\sigma \coloneqq (\theta, i)\\
      \textsc{Substitutions} \quad &\theta \coloneqq (l_i \mapsto t)\dots
    \end{align*}
  \end{minipage}
    \caption{Core grammar of our language}\label{fig:language}
\end{figure}

We assume the existence of several pairwise-disjoint infinite sets: syntactic variables, relation names, logic variables (written $l_n$ which represents the $n^{th}$ logic variable introduced in a given state), and constants.
The full term language consists of the closure of a binary cons constructor \texttt{:} over the union of the syntactic variables, logic variables, and constants.
Our miniKanren language provides a primitive goal $\top$ that always succeeds, and five primitive constructors for term equations, relation calls, binary conjunction, binary disjunction, and variable introduction.

A pair of a goal $G$ and a \emph{state} $\sigma$ is itself a search tree.
A state $\sigma$ is a pair $(\theta,i)$ where $\theta$ is a \textit{substitution} mapping logic variables to terms and $i$ is a numeric index.
To initialize the computation the programmer's \texttt{(run (L) G .\ldots)} query is transformed into an existentially quantified fresh goal $\mathtt{\exists\;(L)\;G\;\ldots}$, and injected into a pair with the \emph{initial state}.
The initial state consists of an empty substitution and 0.

Every leaf node of a search tree is either of the form $G\;\sigma$ or \textsc{empty}, which represents an empty search tree node.
Complex search trees are built from simpler search trees using search tree combinators.
Binary constructors $S\rightarrow S$ and $S\leftarrow S$ represent pointed disjunctions, where the arrow depicts the search ordering of its children.
S + S is an undirected disjunction, and only appears in an answer stream.
S $\times$ G represents a conjunction of a search tree with a goal.
Computation proceeds first down the left child; the goal on the right-hand side does not have a state because its state is determined by the search down the left-hand side.
Special constructors \texttt{delay S} and \texttt{go S} mark the \enquote{delay points} for fair interleaving, points where the engine can suspend/resume goals.

We specify a set of evaluation contexts in \cref{fig:eval-contexts} that identify the next reducible subexpression within a program.
\begin{figure}[ht]
\small
\begin{align*}
  E_P &\coloneqq \texttt{prg}\;\Sigma\;\square\\[4pt]
  E_A &\coloneqq \square \mid (\top\;\sigma)+E_A\\[4pt]
  E_S &\coloneqq \square \mid E_S\leftarrow S \mid S\rightarrow E_S \mid E_S\times G\\[4pt]
  E &\coloneqq E_P[E_A[E_S\;\square]]
\end{align*}
\caption{Evaluation contexts}\label{fig:eval-contexts}
\end{figure}
The program context $E_P$ focuses on the search tree inside of a program; all reductions occur within the search tree.
An answer stream context $E_A$ brings the focus to the search tree just below the sequence of finished answers.
A search context $E_S$ brings the focus to the node within the search tree that is to be reduced.
Intuitively, this context descends into the search tree following the directive of disjunctions and the tree component of conjunctions.
Finally, we define $E$, the composition of the previous three contexts to locate within a program the next search tree node to reduce.

\subsection{Semantics}\label{sec:semantics}
\cref{fig:micro-semantics} presents the small-step reduction rules for our $\mu$Kanren inspired language.
We model the miniKanren search behavior described in \citet{nearlymacrofree}.
In our semantics the program state contains an explicit representation of the entire search tree as it is being explored.
The benefit is that the reduction rules can manipulate this tree structure to simulate goal interleaving (for example, by taking a suspended goal from one branch and moving it to an active position in another branch).
Our semantics explicitly represents individual steps within the execution of each miniKanren goal, such as initiating the evaluation of a goal, suspending it, resuming it, and ultimately completing it (success or failure).
We refer to these individual evaluation points as intra-goal evaluation steps.
Most of the reduction rules in our semantics utilize the $E$ context with the exception of those which match on the relation environment (\textsc{Proceed}) and those which must occur just below the answer stream (\textsc{InvokeDelay, PromoteLeft/Right}).

\begin{figure}[htbp]
\SMALL
\begin{mathpar}
\inferrule[DistrDisj]{}{E[(G_1\vee G_2)\,\sigma] \longrightarrow E[(G_1\,\sigma)\leftarrow(G_2\,\sigma)]}

\inferrule[DistrConj]{}{E[(G_1\wedge G_2)\,\sigma] \longrightarrow E[(G_1\,\sigma)\times G_2]}

\inferrule[LeftAnsConj]{}{E[(\answer\leftarrow S)\times G] \longrightarrow E[(\answer\times G)\leftarrow(S\times G)]}

\inferrule[RightAnsConj]{}{E[(S\rightarrow\answer)\times G] \longrightarrow E[(S\times G)\rightarrow(\answer\times G)]}

\inferrule[AssocRightLeft]{}{E[S_1\rightarrow(\answer\leftarrow S_2)] \longrightarrow E[\answer\leftarrow(S_1\rightarrow S_2)]}

\inferrule[AssocRightRight]{}{E[S_2\rightarrow(S_1\rightarrow\answer)] \longrightarrow E[(S_2\rightarrow S_1)\rightarrow\answer]}

\inferrule[AssocLeftLeft]{}{E[(\answer\leftarrow S_1)\leftarrow S_2] \longrightarrow E[\answer\leftarrow(S_1\leftarrow S_2)]}

\inferrule[AssocLeftRight]{}{E[(S_1\rightarrow\answer)\leftarrow S_2] \longrightarrow E[(S_1\leftarrow S_2)\rightarrow\answer]}

\inferrule[SuccConj]{}{E[\answer\times G] \longrightarrow E[G\,\sigma]}

\inferrule[PruneConj]{}{E[\texttt{empty}\times G] \longrightarrow E[\texttt{empty}]}

\inferrule[PruneLeft]{}{E[\texttt{empty}\leftarrow S] \longrightarrow E[S]}

\inferrule[PruneRight]{}{E[S\rightarrow\texttt{empty}] \longrightarrow E[S]}

\inferrule[SubstFresh]{
  i'= i + |x\dots| \\
  G' = G[x\dots / l_i\dots]
}{
  E[(\exists\;(x\dots)\;G)\;(\theta,i)] \longrightarrow E[G'\;(\theta,i')]
}

\inferrule[Delay]{}{E[r(t_1\dots)\,\sigma] \longrightarrow E[\texttt{delay}\;(\texttt{go}\;r(t_1\dots)\,\sigma)]}

\inferrule[Proceed]{
  G' = G[x\dots/t\dots]
}{
  (prg\;(\dots r(x\dots) \leftrightarrow G\dots)E_A[E_S[\texttt{go}\;r(t\dots)\;\sigma]]) \longrightarrow
  (prg\;(\dots r(x\dots) \leftrightarrow G\dots)E_A[E_S[G'\,\sigma]])
}

\inferrule[UnifySucc]{
  \theta' = \text{unify}(t_1, t_2, \theta)
}{
  E[((t_1 \equiv t_2)\;(\theta ,i))] \longrightarrow E[(\top\;(\theta' ,i))]
}

\inferrule[UnifyFail]{
  \nexists\;\text{mgu}(t_1, t_2, \theta)
}{
  E[((t_1 \equiv t_2)\;(\theta ,i))] \longrightarrow E[\texttt{empty}]
}

\inferrule[DelayConj]{}{E[((\texttt{delay}\;S)\;\times \;G)] \longrightarrow E[(\texttt{delay}\;(S\;\times\;G))]}

\inferrule[DelayLeft]{}{E[((\texttt{delay}\;S_1) \leftarrow S_2)] \longrightarrow E[(\texttt{delay}\;(S_1 \rightarrow\ S_2))]}

\inferrule[DelayRight]{}{E[(S_1 \rightarrow\ (\texttt{delay}\;S_2))] \longrightarrow E[(\texttt{delay}\;(S_1 \leftarrow\ S_2))]}

\inferrule[InvokeDelay]{}{E_P[E_A[(\texttt{delay}\;S)]] \longrightarrow E_P[E_A[S]]}

\inferrule[PromoteLeft]{}{E_P[E_A[(\answer \leftarrow S)]] \longrightarrow E_P[E_A[(\answer + S)]]}

\inferrule[PromoteRight]{}{E_P[E_A[(S \rightarrow \answer)]] \longrightarrow E_P[E_A[(\answer + S)]]}

\end{mathpar}
\caption{microKanren language semantics}\label{fig:micro-semantics}
\end{figure}

We limit the amount of interleaving by introducing delay points only at relation invocations (\textsc{Delay}).
This results in a more biased, less interleaved search than \citeauthor{dmitrysemantics}'s~\cite{dmitrysemantics} model.
\Cref{fig:minimal-example} shows a trace demonstrating how delay and forcing of delays behaves on a small example.

\begin{center}
\small
\begin{minipage}{0.95\linewidth}
\begin{align*}
 & \quad \prg{\texttt{same}(x,y) \leftrightarrow x \equiv y}{(\exists (p)\; \texttt{same}(p,\strcat))\; (\emptyset, 0)} \\[0.3em]
\step{\text{\textsc{SubstFresh}}}
& \quad \prg{\texttt{same}(x,y) \leftrightarrow x \equiv y}{\texttt{same}(0,\strcat)\; (\emptyset, 1)} \\[0.3em]
\step{\text{\textsc{Delay}}}
& \quad \prg{\texttt{same}(x,y) \leftrightarrow x \equiv y}{\delay(\go(\texttt{same}(0,\strcat)))\; (\emptyset, 1)} \\[0.3em]
\step{\textsc{InvokeDelay}}
& \quad \prg{\texttt{same}(x,y) \leftrightarrow x \equiv y}{\go(\texttt{same}(0,\strcat))\; (\emptyset, 1)} \\[0.3em]
\step{\textsc{Proceed}}
& \quad \prg{\texttt{same}(x,y) \leftrightarrow x \equiv y}{0 \equiv \strcat\;(\emptyset, 1)} \\[0.3em]
\step{\textsc{UnifySucc}}
& \quad \prg{\texttt{same}(x,y) \leftrightarrow x \equiv y}{\top\; ((0 \mapsto \strcat), 1)}
\end{align*}
\captionof{figure}{Worked example trace: evaluating $(\exists (p)\; \mathtt{same}(p,\strcat))$ under our semantics.}\label{fig:minimal-example}
\end{minipage}
\end{center}

To show how the interleaving and resumption machinery behaves, we now sketch a trace for a query with a disjunction.
This example will show how fair interleaving is implemented by step-wise promotion and resume.
Delayed subtrees are explicitly wrapped with a \texttt{delay} tag and bubbled upward where they interleave disjunctions (\textsc{DelayLeft/Right}) and bypass conjunctions (\textsc{DelayConj}).
Once the delay tag reaches the tail of the answer stream, it is released (\textsc{InvokeDelay}) and evaluation continues in the freshly interleaved tree.
\Cref{fig:conde-promotion-trace} shows key steps in evaluating a disjunctive query with two \texttt{same} calls, illustrating how our semantics models promotion and fair interleaving of branches.

\begin{center}
\small
\begin{minipage}{0.95\linewidth}
\begin{align*}
 & \quad \prg{\texttt{same}(x,y) \leftrightarrow x \equiv y}{(\texttt{same}(0,\strcat) \vee \texttt{same}(0,\strdog))\; (\emptyset, 1)} \\[0.3em]
\step{\textsc{DistrDisj}}
 & \quad \prg{\texttt{same}(x,y) \leftrightarrow x \equiv y}{(\texttt{same}(0,\strcat)(\emptyset,1) \leftarrow \texttt{same}(0,\strdog)(\emptyset,1)} \\[0.3em]
\step{\textsc{Delay} + \textsc{Go (Left branch)}}
 & \quad \prg{\texttt{same}(x,y) \leftrightarrow x \equiv y}{\delay(\go(\texttt{same}(0,\strcat))) \leftarrow \texttt{same}(0,\strdog)\; (\emptyset, 1)} \\[0.3em]
\step{\textsc{Proceed + UnifySucc}}
 & \quad \prg{\texttt{same}(x,y) \leftrightarrow x \equiv y}{\top\;((0 \mapsto \strcat), 1) \leftarrow \texttt{same}(0,\strdog)\; (\emptyset, 1)} \\[0.3em]
\step{\textsc{PromoteRight}}
 & \quad \prg{\texttt{same}(x,y) \leftrightarrow x \equiv y}{(\top\;((0 \mapsto \strcat), 1) + \texttt{same}(0,\strdog))\; (\emptyset,1)}
\end{align*}
\captionof{figure}{Trace for \texttt{(conde ((same q 'cat)) ((same q 'dog)))}, showing interleaving promotion.}\label{fig:conde-promotion-trace}
\end{minipage}
\end{center}

\subsection{Novel design choices}

Our small-step semantics were designed with visual debugging in mind, which influenced many decisions regarding the language.
\paragraph{Explicit Answer Stream} Successful goals are not immediately removed from the search unlike that of \citet{dmitrysemantics}.
    Instead, a success is represented as $(\top\, \sigma)$ and remains as part of the structure until it is lifted to the top-level answer stream (\textsc{PromoteLeft/Right}).
    We model the answer stream as a sequence of these answer nodes accumulated with the $+$ operator.
    This allows the user to inspect each answer as it is produced and to see it in context.
    It also preserves a record of answers which can be useful for understanding the progress of the search.
    While success is explicitly modeled, failure (represented as \texttt{empty}) is not.
    Upon unification failing (\textsc{UnifyFail}), that subtree is promptly pruned from the tree (\textsc{PruneConj, PruneLeftRight}) mimicking the silent refutation of miniKanren.
\paragraph{Pointed Disjunction (Railway Model)} We represent disjunctions in a symmetric, bidirectional way to preserve the relative orderings of disjuncts upon interleaving (\textsc{DelayLeft/Right}).
    We refer to this as the \textit{railway model} as the two branches of a disjunction are like a switch in railway tracks where evaluation can proceed down either side.
    In the grammar, these are represented as $\leftarrow$ and $\rightarrow$.
    Traditionally, a singular left-biased disjunction is used and interleaving involves swapping the two disjuncts.
    Visually and textually, this makes it more difficult to retain information about the search tree and requires the user to recalibrate to its new shape.
    While changing a $\leftarrow$ to a $\rightarrow$ or vice versa is a subtle syntactic shift, it promotes better structural correspondence between successive reductions---particularly when evaluation contexts are made explicit.
    As seen in \cref{fig:micro-semantics}, this combinatorially affects the number of certain reduction steps (\textsc{AssocLeftLeft, etc., PromoteLeft/Right, Left/RightAnsConj, DelayLeft/Right, PruneLeft/Right}); however, a correctness preserving yet smaller semantics could be derived by collapsing the bidirectional disjunctions into a single unoriented disjunction operator.
\paragraph{Delayed Relation Expansion} In addition to the \texttt{delay} operator, we also use a special \texttt{go} operator that wraps relation invocations.
    This thunk-like form prevents the immediate expansion of a relations body (\textsc{Delay}) until it is scheduled for evaluation (\textsc{Proceed}).
    By delaying unnecessary expansion, we avoid exploding the search tree too early, making the focus on the current evaluation less cluttered while maintaining a more intuitive conceptualization of the actual goal being suspended (i.e., a relation call with certain arguments rather than whatever its goal body may be).
The way that we minimized the expansion of relcalls until we actually use them (see \cref{sec:visu-reduct} for an example).
Further debugging-related extensions, including state and goal-level source mapping and additional state information, are described in \cref{sec:mixed-redex-js}.

\subsubsection{Semantics validation:} Using PLT Redex, we tested two key properties of our reduction semantics.
We defined a well-formedness predicate on our search states (ensuring, for example, that paused goals carry a valid state, variables are fresh where expected, etc.).
Using that, we were able to test the following:
\begin{itemize}
\item \textbf{Determinism:} Starting from any well-formed initial configuration, the reduction relation has at most one possible next step.
  In other words, the semantics is a deterministic small-step interpreter for miniKanren---there are no ambiguous rule overlaps or nondeterministic choices.
  We validated this by random generation of states and checking that Redex never found two distinct reductions for the same state.
\item \textbf{Syntactic Preservation}
  Using Redex's random testing, we confirmed that every reduction step preserves this invariant---only well-formed states produce well-formed successor states.
  This gives confidence that the formal rules faithfully maintain the intended structure of the search tree, and do not lead to malformed or inconsistent states as the search proceeds.
\end{itemize}

While well short of a formal proof, these property tests serve as valuable sanity checks.
Together, the Redex validation suggests that our semantics is sound and behaves as expected for a wide range of cases.

\subsubsection{Alternative interleave strategies}

Our semantics implement interleaved exploration of different branches of the search tree.
As long as some delay will occur along any cycle in the call graph, no single disjunct or recursive path monopolizes computation, and no potential solution is starved indefinitely.
This guarantees completeness even in the presence of infinite branches.

Various miniKanren implementations differ in where they place delay points.
These choices directly impact the search order, and so impact queries' efficiency.
%
%
A key benefit of reduction semantics is their modularity~\cite{felleisen1992revised}.
In our case, this means we can easily swap out reduction relations to explore different search strategies.

To that end, we implement an alternative semantics that uses Prolog's depth-first search within the same framework.
It uses an interleaving depth-first search strategy to enumerate substitutions that satisfy a given goal.
In contrast with miniKanren's interleaving search, the depth-first search (DFS) of Prolog is more efficient but incomplete.
By interleaving nowhere, we can model the DFS search semantics, which serves as a baseline for comparison and to illustrate the ease of modeling alternative strategies in our model.

\section{A Mixed Redex-JS Visualizer}\label{sec:mixed-redex-js}

We provide a web-based interface with a custom tree visualization that utilizes the executable semantics in our Redex model.
By visualizing the search in tree form, the user can see the entire state of the search and observe the operational search behavior including the branching of goals, where goals are suspended or resumed, how execution proceeds incrementally, how state is propagated and expanded, and more.
While visual tools of this kind are not inherently better than textual ones~\cite{patel1994texttree}, the tree structure of our visualizer is more faithful to the nature of the computation and information can be presented in a more comprehensible manner.
We paid special attention to addressing usability issues noted in studies of earlier logic-program visualizers (~\cref{sec:related-work}).
For example, our interface explicitly marks the current point of execution (solving the common problem of \enquote{unclear current goal} in static tree diagrams) and clearly displays the variable bindings in the current state (making the construction of output substitutions transparent).

Primarily, our visualizer serves as a pedagogical stepping visualizer---a "notional machine" that helps students precisely understand how miniKanren executes logic programs step by step. 
While the tool provides insight into the evaluation process and search order, it is not a full-fledged debugger with advanced breakpoint or query capabilities. 
Nonetheless, experienced users might employ it as a limited diagnostic aid for manually tracing and understanding surprising search behaviors.
In this section we describe the visualizer including additional infrastructure beyond the Redex model, augmentations to our language and semantics, as well as components of the UI front-end.

\subsection{Architecture}\label{sec:architecture}
\begin{figure}[hbt]
    \centering
    \includegraphics[width=\textwidth]{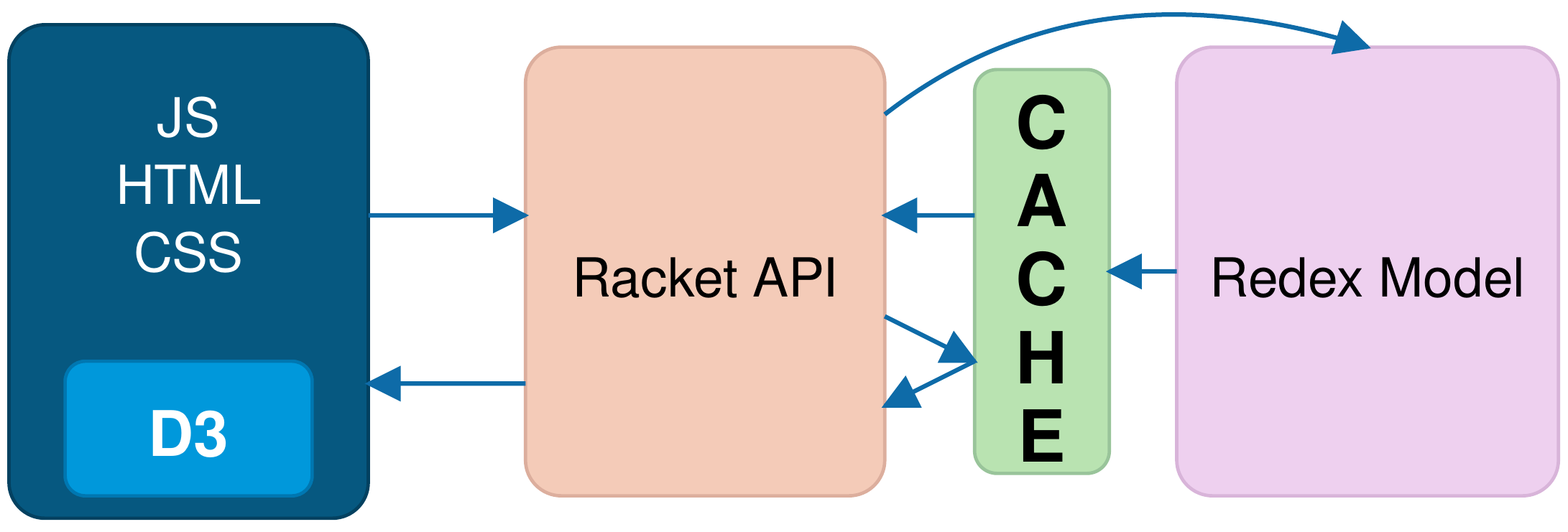}
    \caption{The architecture of our visualizer.}\label{fig:architecture}
\end{figure}

\cref{fig:architecture} shows the high-level architecture of visualizer.
A User requests such as submission of a program to be visualized, stepping forwards or backwards, resetting, or changing semantics is sent from the frontend to the backend API.
Users begin by selecting a set of reductions (currently either $\mu$Kanren and depth-first).
The user then submits a program written in the concrete syntax of \citet{reasoned2}
Having a canonical and sugared surface syntax makes this tool accessible for a wide audience without having to learn a new syntax or rewrite existing programs.
Additionally, users can explore different semantics over the same exact program base.
Upon receiving a new program, a check leveraging \texttt{hosted-minikanren} and the \texttt{syntax-spec} frontend~\cite{syntaxspec} supplemented by a custom well-formed judgment is performed to detect static errors.
This will capture errors such as illegal expressions, parentheses mismatches, relation arity mistakes, and binding errors.
If there is such an error, it is reported back to the user and visualization will not proceed until a valid program is entered.
Once a valid program is entered, it is transpiled to the language of our model.
At this point, goals and the initial state are uniquely tagged to facilitate tracing (see \cref{sec:lang-improv}).

Once a program has been loaded, execution proceeds incrementally: each step forward applies a single reduction rule from the model, serializes the updated search tree as JSON, and transmits it to the front end for rendering via the $D^3$ library \cite{d3}.
We maintain a list with a zipper~\cite{huet1997zipper} as a cache of previous states allowing users to step backwards (without a cache this would not be possible) and forwards through the history.
This allows us constant time access for stepping forwards and backwards through the cache as well as insertion at the cost of linear space.
If the forward cache is empty and the user requests to step forward, the current program is stepped and cached while the new program gets sent to the front end.

\subsection{Language Improvements}\label{sec:lang-improv}
The language and semantics described in \cref{sec:language-semantics} provides a minimal core of interleaving search which we augment with additional information to aid the visualizing experience.
This methodology is similar to other examples in the literature such as \citet{bernstein1995debugsemantics} and \citet{kamburjan2021debugsemantics}.
It should be noted that these changes result in an equivalent semantics.
\textbf{Goal and State Tagging}.
As mentioned in \cref{sec:architecture}, goals and states are given UIDs at transpilation time.
For goals, this allows for users to bidirectionally trace between the source code they provided and the dynamic tree visualization.
This is useful for locating easily locating problematic goals or for understanding where execution is occurring.
For states, this enables "subscribing" to a state and following it as it proceeds through execution. 
Taking unification as an example, the previous non-terminal $t_1 \equiv t_2$ becomes $t_1 \equiv t_2 \,c$ where $c$ is the UID.
Other goals follow similarly except for $\top$ which remains untagged as it has no mapping to the source code.
The state non-terminal now becomes $(\theta,i,c)$.
The reduction rules that need modified are those dealing with goal-state evaluations and \textsc{DistrDisj} which must also generate a new UID for the state of its right disjunct.
\textbf{Trail}.
We once again extend the state terminal with what we refer to as a \textit{trail}.
This means a state is now a $(\theta,i,\tau,c)$ where $\tau = (t_1, t_2, c) \dots$ .
A trail is like a substitution except it is always extended by a successful unification and is linearly ordered.
This effectively keeps a record of all the atomic computations that have been applied on a given state.
As the trail retains the UIDs of the unifications, elements of the trail can also be traced back to the source code.
To support this, the \textsc{UnifySucc} rule must append the triple of the walked terms and the UID to the current accumulated trail.
\textbf{Reification}.
Although not an extension of the language, we also reify our substitutions before sending them to the visualizer.
To do so, we construct a \texttt{run*} with the set of initial query variables and a \texttt{fresh} for the logic variables created at runtime.
Inside the fresh, unifications of the mappings of the substitution are conjuncted.
%
This can result in at most one answer as each substitution represents at most one answer.
This is obviously useful for nodes in the answer stream, but we also perform reification on states with more work to be done.
This can be useful as the reified representation can reveal information about constraints on the solution such as "this solution must be a pair, although the elements of that pair is not yet known" or "this solution must be 'dog and anything that comes along and contradicts that will cause this branch to fail".
\subsection{Visualizer UI}\label{vis-ui}

\begin{figure}[htb]
  \centering
  \fbox{\includegraphics[width=0.8\textwidth]{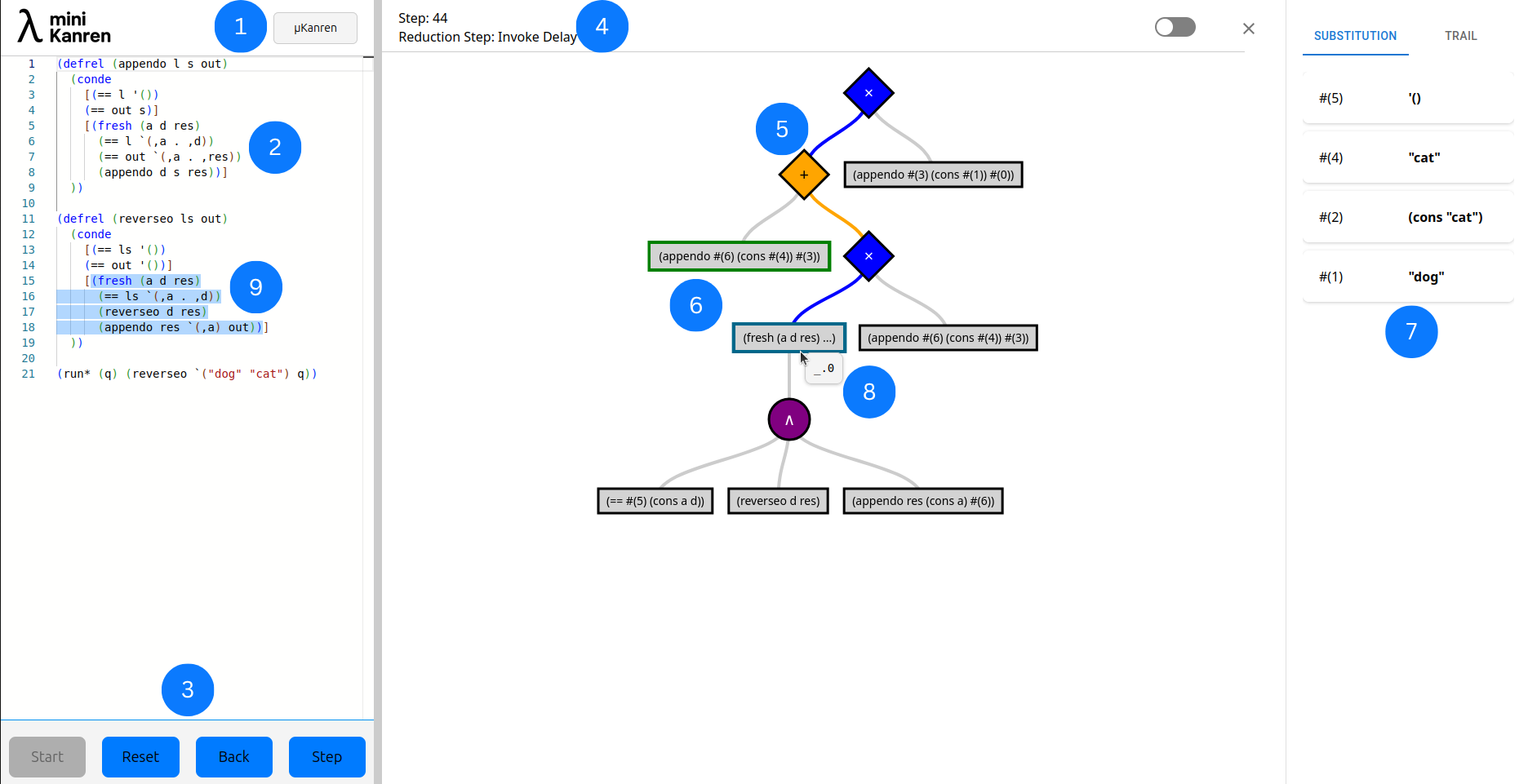}}
  \caption{The visualizer.}\label{fig:viz}
\end{figure}

\begin{figure}[htbp]
  \centering
  \renewcommand{\arraystretch}{1.5}
  \begin{tabular}{@{} l @{\hspace{1em}} l @{}}
    \raisebox{-0.5\height}{\includegraphics[height=1.2cm]{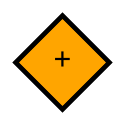}}     & \textbf{Tree Disjunction} \\
    \raisebox{-0.5\height}{\includegraphics[height=1.2cm]{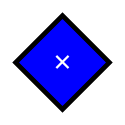}}      & \textbf{Tree Conjunction} \\
    \raisebox{-0.5\height}{\includegraphics[height=1.2cm]{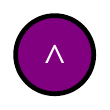}}     & \textbf{Goal Disjunction} \\
    \raisebox{-0.5\height}{\includegraphics[height=1.2cm]{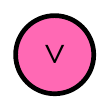}}      & \textbf{Goal Conjunction} \\
    \raisebox{-0.5\height}{\includegraphics[height=1.2cm]{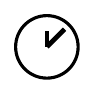}}    & \textbf{Delay} \\
    \raisebox{-0.5\height}{\hspace{-1cm}\includegraphics[height=1cm]{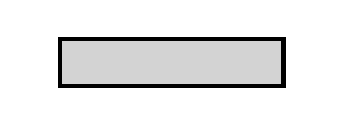}} & \textbf{Text Node} (Relation Call, Fresh, Unification) \\
    \raisebox{-0.5\height}{\hspace{0.1cm}\includegraphics[height=1cm]{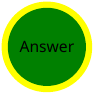}}   & \textbf{Answer} \\
    \raisebox{-0.5\height}{\hspace{-0.1cm}\includegraphics[height=1.4cm]{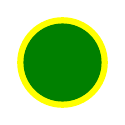}}  & \textbf{Success} \\
    \raisebox{-0.5\height}{\includegraphics[height=1.2cm]{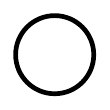}}    & \textbf{Failure} \\
  \end{tabular}

  \vspace{0.5em}
  \small
  \textbf{Additional highlighting:} Yellow = \textit{State}, Green = \textit{Marked Go}, Blue = \textit{Selected By User}
  \caption{Node shape legend used in the visualizer.}\label{fig:node-table}
\end{figure}

\cref{fig:viz} shows the interface for the visualizer.
The user begins by selecting a set of reductions (1) and entering their program (2).
Once the program is entered, the user can hit the \textit{start} button and begin stepping through execution (3).
At every evaluation step, the visualizer presents a rich snapshot of program state.
Above the tree is displayed the current step count and the name of the last applied reduction rule (4).
The evaluation context is visible by means of tracing the colored edges down to a subtree (5). 
Here, you can also see the railway model as the disjunction node in orange is right-facing.
Additionally, stateful nodes---those which are highlighted (6)---are interactive: clicking on such a node reveals a sidebar with the current substitution and trail (7). 
Its reified form with respect to the original query variables can be seen by hovering over such a node (8).
Finally, goals can be traced bidirectionally from the source code to the tree visualization by simply clicking on them (9).
Clicking on a goal in the source code will highlight all occurrences of that goal (if any) in the tree, and clicking on a goal in the tree will highlight the corresponding goal in the source.
A full description of the nodes and their meanings is available in \cref{fig:node-table}.

\section{Visualizing Reductions}\label{sec:visu-reduct}



As discussed in \cref{sec:introduction}, understanding why programs produce answers in specific orders is crucial for reasoning about logic programs effectively. 
In this section, we demonstrate how to leverage our custom visualizer to inspect and understand search behaviors clearly. 
By stepping through evaluation with both depth-first and interleaving semantics, users can directly observe how different operational semantics influence the order of answers generated by a query. 
Through illustrative examples, we showcase how this visualization aids users in understanding search scheduling, debugging faulty implementations, and optimizing performance by identifying differences in goal evaluation order.

\subsection{Using the visualizer}\label{sec:visu-reduct-1}

\paragraph{Understanding Search Behavior}

%
%
As promised, we will explore why the program in \cref{sec:introduction} produces answers in the order it does.

\begin{figure}[htb]
\centering
\vspace{-0.5cm}
\begin{tikzpicture}[
  node distance=0.5cm and 0.5cm,
  imagebox/.style={inner sep=0, anchor=center},
  every node/.style={anchor=center}
]

\node[imagebox] (img1) at (0, 0) {\includegraphics[width=3.5cm, height=4cm, keepaspectratio]{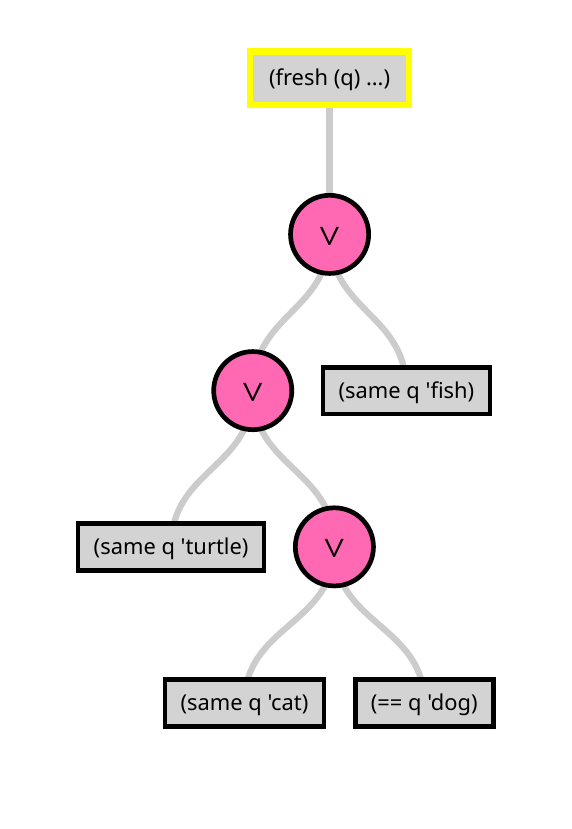}};
\node[imagebox, right=of img1] (img2) {\includegraphics[width=3.5cm, height=4cm, keepaspectratio]{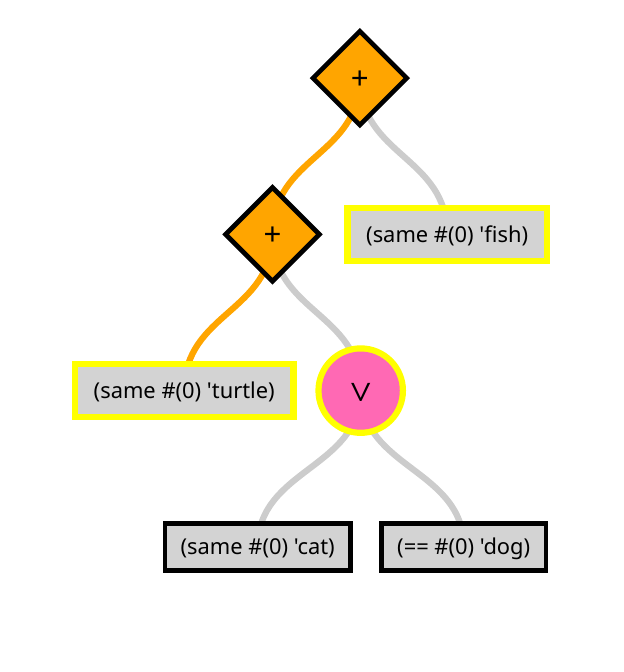}};
\node[imagebox, right=of img2] (img3) {\includegraphics[width=3.5cm, height=4cm, keepaspectratio]{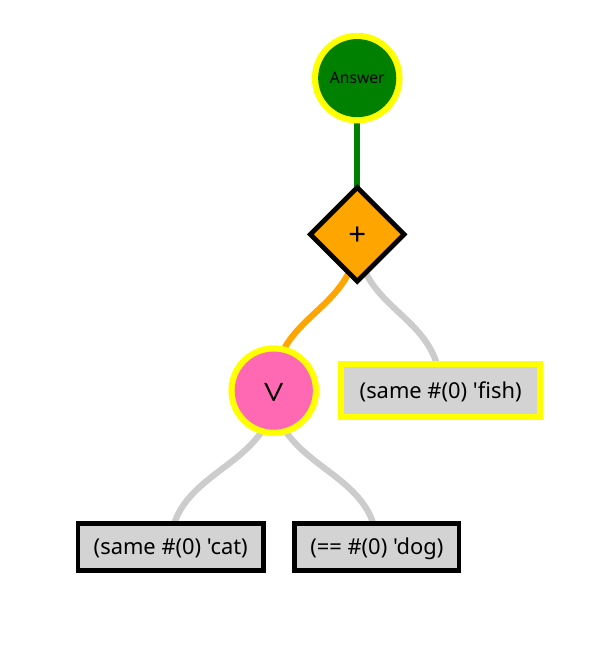}};

\node[imagebox, below=of img1] (img4) {\includegraphics[width=3.5cm, height=4cm, keepaspectratio]{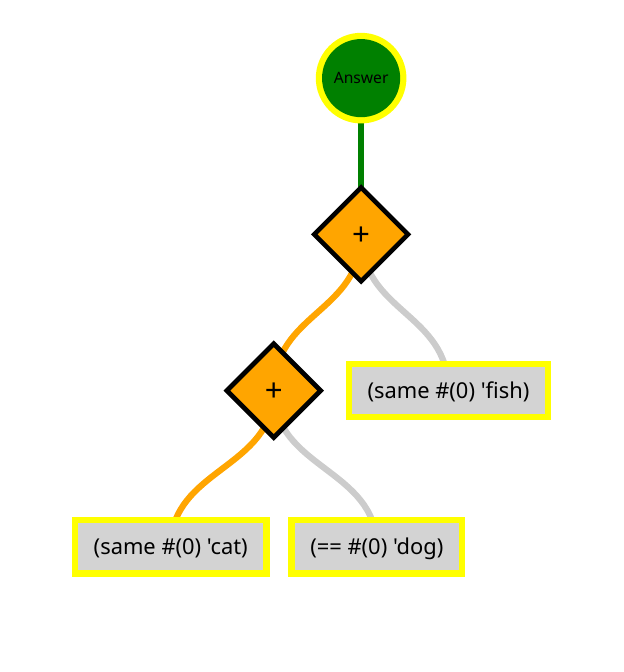}};
\node[imagebox, right=of img4] (img5) {\includegraphics[width=3.5cm, height=4cm, keepaspectratio]{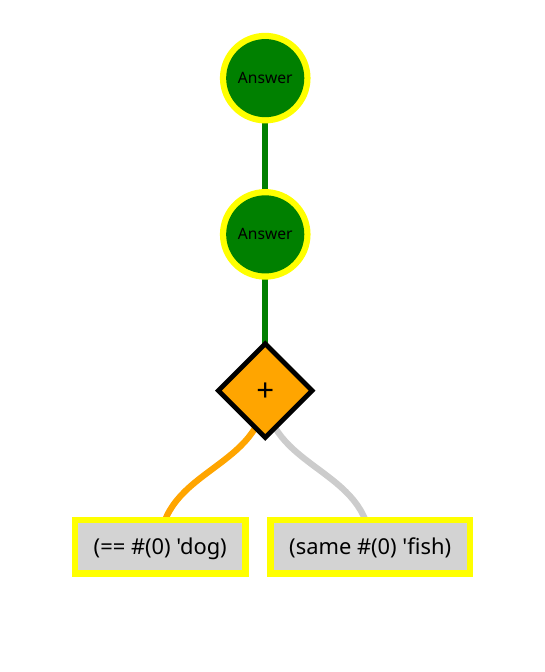}};
\node[imagebox, right=of img5] (img6) {\includegraphics[width=3.5cm, height=4cm, keepaspectratio]{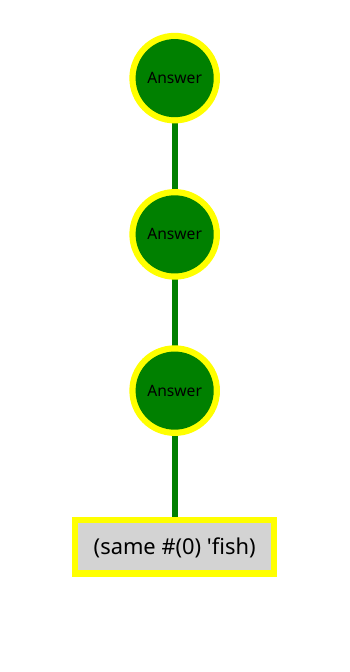}};
\node[imagebox, right=of img6] (img7) {\includegraphics[width=3.5cm, height=4cm, keepaspectratio]{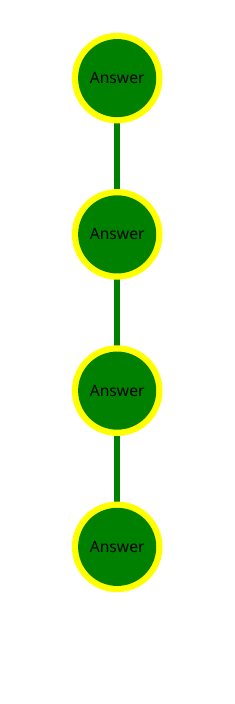}};

\draw[->, thick] (img1.east) -- (img2.west) node[midway, above] {\small x3};
\draw[->, thick] (img2.east) -- (img3.west) node[midway, above] {\small x4};
\draw[->, thick] (img3.south) -- ++(0,-0.35) -- ++(-7.35,0) node[midway, above, yshift=2pt] {\small x1} -- ++(0,-0.4);
\draw[->, thick] (img4.east) -- (img5.west) node[midway, above] {\small x4};
\draw[->, thick] (img5.east) -- (img6.west) node[midway, above] {\small x2};
\draw[->, thick] (img6.east) -- (img7.west) node[midway, above] {\small x2};

\end{tikzpicture}
\caption{Evaluation showing DFS behavior.}\label{fig:example-dfs}
\end{figure}
First, we will explore the execution with the depth-first search semantics which enumerates answers in a more expected manner.
\cref{fig:example-dfs} shows the execution of the program with arrows indicating the number of intermediate steps that took place.
We begin with a top level fresh query with disjunctions in its body.
The initial state is distributed over the disjunctions and the evaluation context proceeds down the left side until a relation call is reached.
This call, \texttt{(sameo \#(0) 'turtle)}, is the first to succeed and is brought to the top of the answer stream.
Evaluation continues down the left side, distributing the state over another disjunction and encountering \texttt{(sameo \#(0) 'cat)} which succeeds and is lifted into the answer stream.
Finally, the goal \texttt{(== \#(0) 'dog)} succeeds and is promoted to the answer stream and \texttt{(sameo \#(0) 'fish)} follows.
With this example, the search is straightforward as it is a pure depth-first search with no interleaving.
The order of the answers is the same as the order in which the clauses were written.

\begin{figure}[htb]
\centering
\vspace{-0.5cm}
\begin{tikzpicture}[
  node distance=0.5cm and 0.5cm,
  imagebox/.style={inner sep=0, anchor=center},
  every node/.style={anchor=center}
]

\node[imagebox] (img1) at (0, 0) {\includegraphics[width=3.5cm, height=4cm, keepaspectratio]{picts/ex1_1.pdf}};
\node[imagebox, right=of img1] (img2) {\includegraphics[width=3.5cm, height=4cm, keepaspectratio]{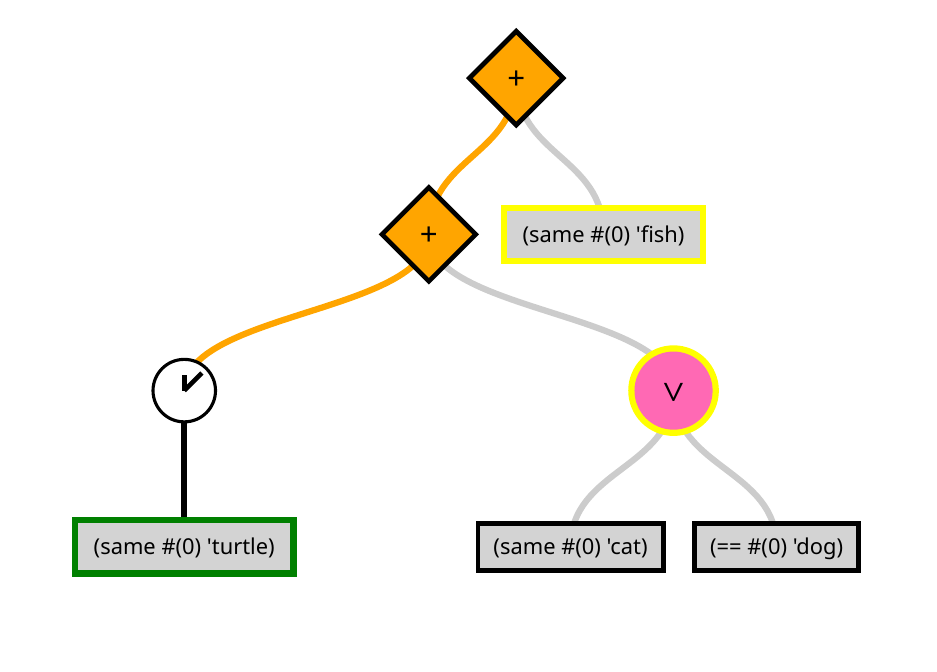}};
\node[imagebox, right=of img2] (img3) {\includegraphics[width=3.5cm, height=4cm, keepaspectratio]{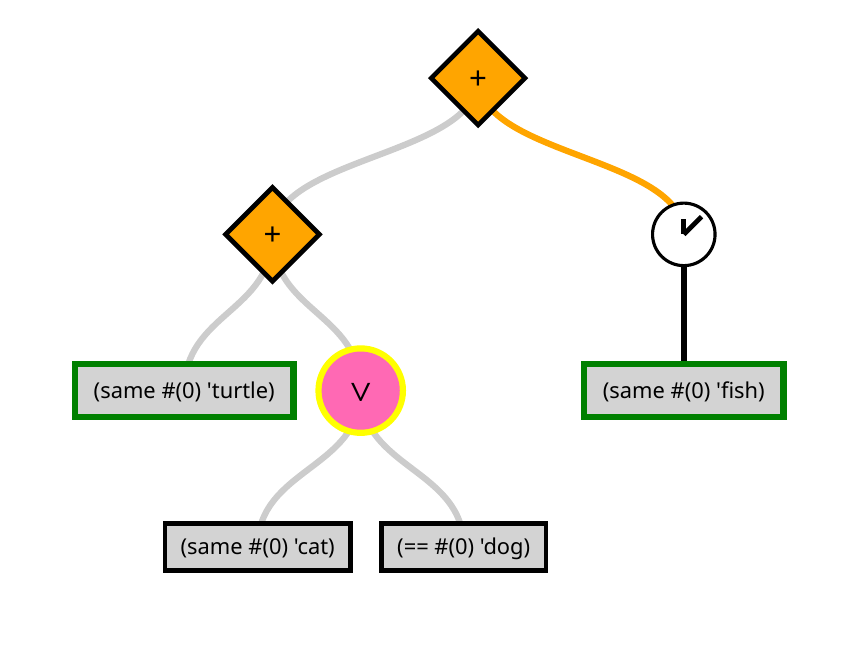}};

\node[imagebox, below=of img1] (img4) {\includegraphics[width=3.5cm, height=4cm, keepaspectratio]{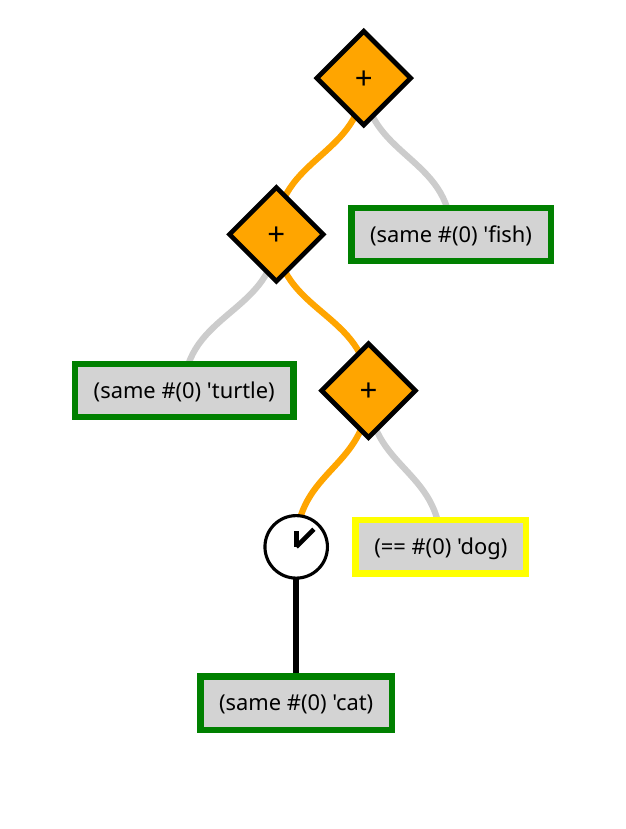}};
\node[imagebox, right=of img4] (img5) {\includegraphics[width=3.5cm, height=4cm, keepaspectratio]{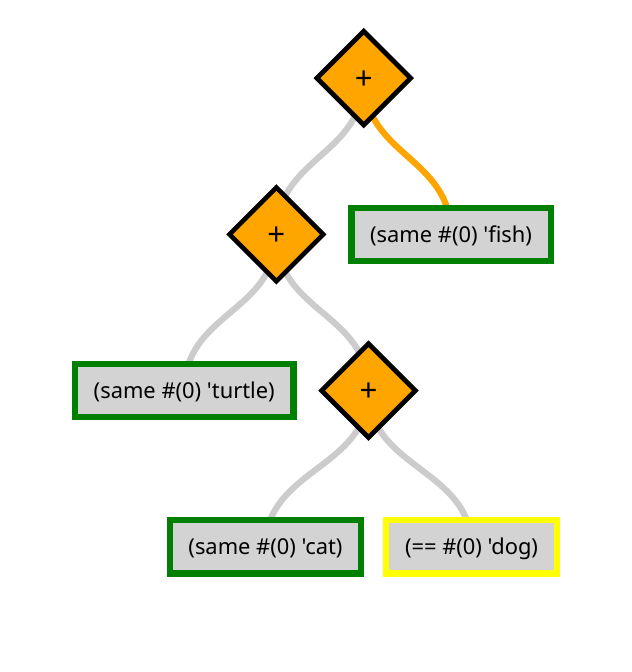}};
\node[imagebox, right=of img5] (img6) {\includegraphics[width=3.5cm, height=4cm, keepaspectratio]{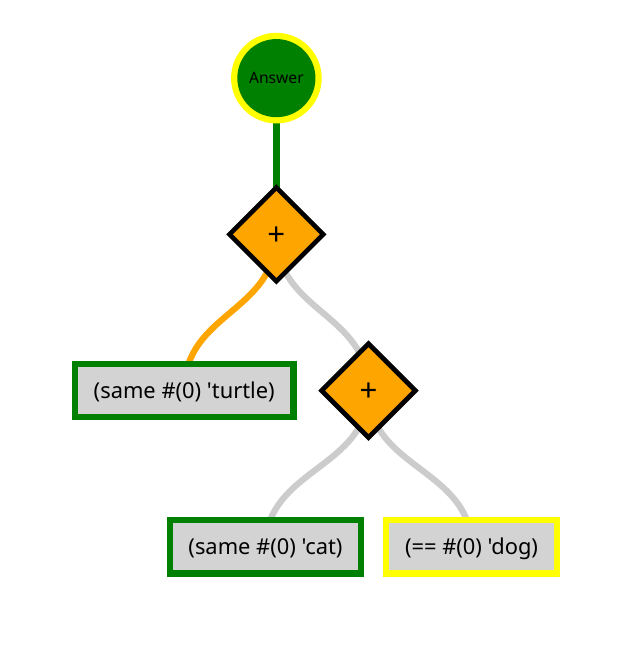}};

\node[imagebox, below=of img4] (img7) {\includegraphics[width=3.5cm, height=4cm, keepaspectratio]{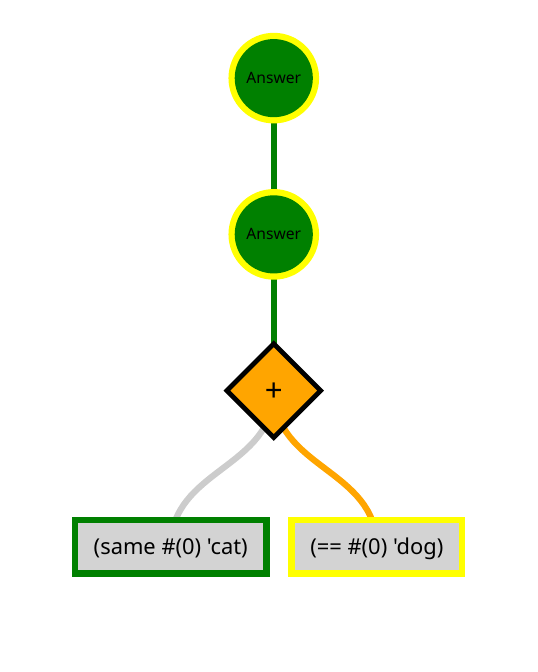}};
\node[imagebox, right=of img7] (img8) {\includegraphics[width=3.5cm, height=4cm, keepaspectratio]{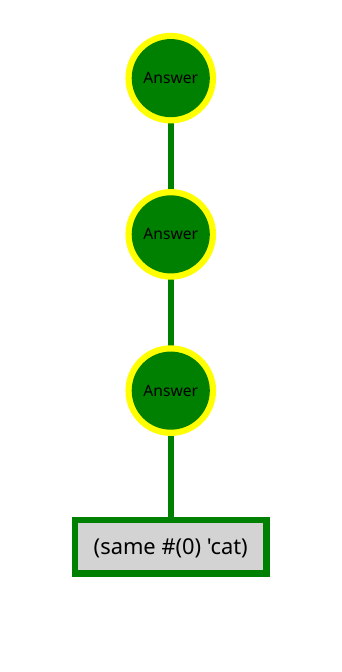}};
\node[imagebox, right=of img8] (img9) {\includegraphics[width=3.5cm, height=4cm, keepaspectratio]{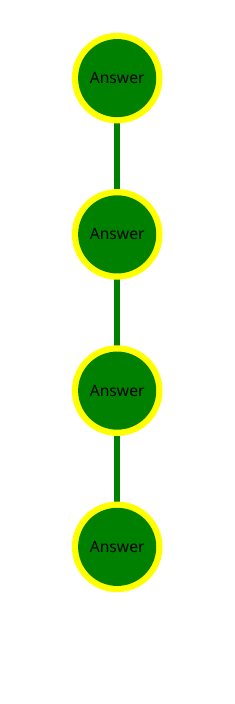}};

\draw[->, thick] (img1.east) -- (img2.west) node[midway, above] {\small x4};
\draw[->, thick] (img2.east) -- (img3.west) node[midway, above] {\small x4};
\draw[->, thick] (img3.south) -- ++(0,-0.5) -- ++(-7.5,0) node[midway, above, yshift=2pt] {\small x4} -- ++(0,-0.5);
\draw[->, thick] (img4.east) -- (img5.west) node[midway, above] {\small x4};
\draw[->, thick] (img5.east) -- (img6.west) node[midway, above] {\small x3};
\draw[->, thick] (img6.south) -- ++(0,-0.3) -- ++(-7.8,0) node[midway, above, yshift=2pt] {\small x3} -- ++(0,-0.3);
\draw[->, thick] (img7.east) -- (img8.west) node[midway, above] {\small x2};
\draw[->, thick] (img8.east) -- (img9.west) node[midway, above] {\small x2};

\end{tikzpicture}
\caption{Evaluation showing interleaving behavior.}\label{fig:example-inter}
\end{figure}
\cref{fig:example-inter} shows the execution of the same program except with interleaving semantics.
The first divergence between these semantics and the depth-first one occurs after four steps: we have substituted our query variable (now \texttt{\#(0)}), distributed the initial state over two conjunctions, but now the first relation call is delayed.
This delay is propagated up, interleaving disjunctions as it bubbles up, until evaluation proceeds down the other side and again delays a relation call.
One can already see the railway model and lazy relation expansions at this point.
Although interleaving has occurred, the ordering of the disjuncts remains consistent.
Additionally, the goals highlighted in green and marked as ready to proceed by the scheduler have not been expanded.
Again, the delay is propagated up and evaluation now proceeds through two disjunctions to yet another relation call.
The process repeats and evaluation is now focused on \texttt{(sameo \#(0) 'fish)} which will succeed and produce the first answer, \texttt{'fish}.
Evaluation continues down the left side of the tree where \texttt{(sameo \#(0) 'turtle)} is encountered and produces the second answer, \texttt{'turtle}.
Next, the right side of the final disjunction, \texttt{(== \#(0) 'dog)}, immediately succeeds without delay since it is not a relation call and yields the third answer, \texttt{'fish}.
Finally, the last goal, \texttt{(sameo \#(0) 'cat)}, succeeds and produces the final answer, \texttt{'cat}.

\paragraph{Debugging a Broken Program}
As the story goes~\cite{rosenblatt2019firstorder}, there was once a novice miniKanren programmer named Nub Let who wrote the following program known by locals as \textit{appendoh the deficient}.
\begin{verbatim}
(defrel (appendoh l s ls)
  (conde
   ((== '() l) (== s ls))
   ((fresh (a d res)
       (== `(,a . ,d) l)
       (== `(,a . ,res) ls)
       (appendoh d s ls)))))
\end{verbatim}
As a sanity check, Nub Let ran the query \texttt{(run* (q) (appendoh '(dog) q '(dog cat)))} but they got out the answer \texttt{'((dog cat))}!
Clearly, something went wrong, so Nub Let threw the program into our visualizer to find the issue.
Nub Let quickly steps the program until they see the first answer appear in the answer stream.
They hover over the node and see the same answer they got before.
\\
\begin{center}
\includegraphics[width=4.5cm]{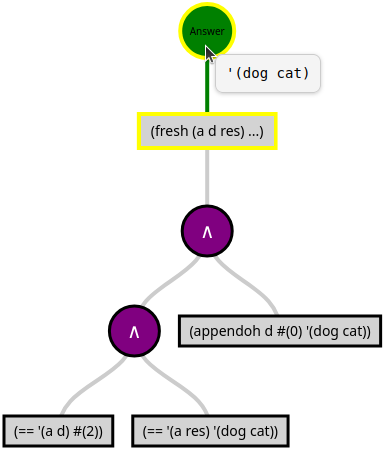}
\end{center}

\noindent Nub Let decides to inspect the trail to see if there are any clues there.
They see that the first argument they provided, \texttt{'(dog)}, was unified with \texttt{(cons \#(1) \#(2))} and similarly the third argument they provided, \texttt{'(dog cat)}, was unified with \texttt{(cons \#(1) \#(3))}.
Nub Let maps these to the source program and sees they occurred in the first and second goals within the \texttt{fresh}, respectively.
Nub Let also observes that \#(2) and the empty list were unified along with \#(0) and \texttt{'(dog cat)} within the first clause of the \texttt{conde}.
Nub Let is not experienced enough to be suspicious that \#(3) corresponding to the fresh variable \texttt{res} introduced in the \texttt{fresh} does not appear in this last unification.
Nub Let decides to subscribe to this state by clicking on the answer node and steps backwards looking for the bug.
After stepping back a few steps, they observe that they are decomposing the first and third arguments into their head and tails and then recursively calling the \texttt{appendoh} relation.

\begin{center}
\includegraphics[width=5.5cm]{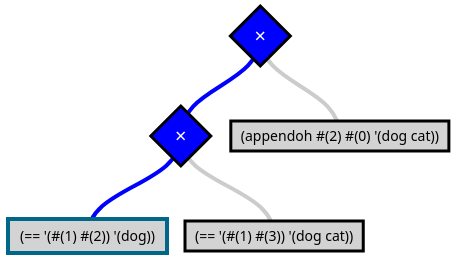}
\end{center}

\noindent Here, Nub Let notices that while the first argument is being shrunk in the recursion, the third element is the same as it was in the query!
They step past the two unifications and observe the substitution.
They see \#(3)---corresponding to the fresh variable \texttt{res}---is mapped to \texttt{'cat} and that this is what they should be recurring with.
They click on the \texttt{appendoh} node, locate it in their source program, and finally change the \texttt{ls} to be \texttt{res}.
\paragraph{Optimizing Performance}
Nub Let has a cousin named Newb Let who is a seasoned logician and who knows that conjunction is commutative. 
Like his cousin, Newb Let is adventuring into logic programming and decides to write his own implementation of \texttt{appendo}.
They figure it should not matter where the recursive call is placed as the order of conjuncts should not matter and the interleaving search will ensure completeness.
\begin{verbatim}
(defrel (appendoh l s ls)
  (conde
   ((== '() l) (== s ls))
   ((fresh (a d res)
       (appendoh d s res)
       (== `(,a . ,d) l)
       (== `(,a . ,res) ls)))))
\end{verbatim}
However, after benchmarking their implementation, Newb Let finds theirs to be deficient as well!
Besides not terminating in some modes, Newb Let's implementation is dramatically slower than the canonical implementation in other modes.
They decide to query their relation with all arguments fresh to diagnose the performance bottleneck.
At first, Newb Let observes an answer appears very quickly originating from the first clause.

\begin{center}
  \includegraphics[width=5cm]{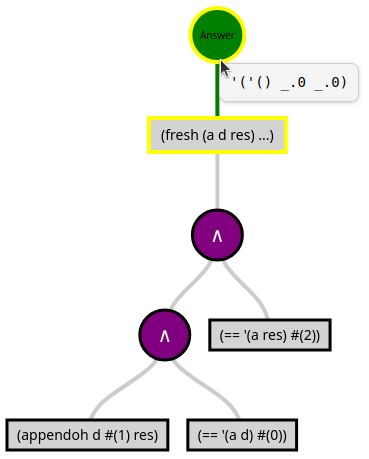}
\end{center}

\noindent They are satisfied with this very general answer and proceed with stepping.
After more stepping, the first recursive call to \texttt{appendoh} is reached.

\begin{center}
  \includegraphics[width=5cm]{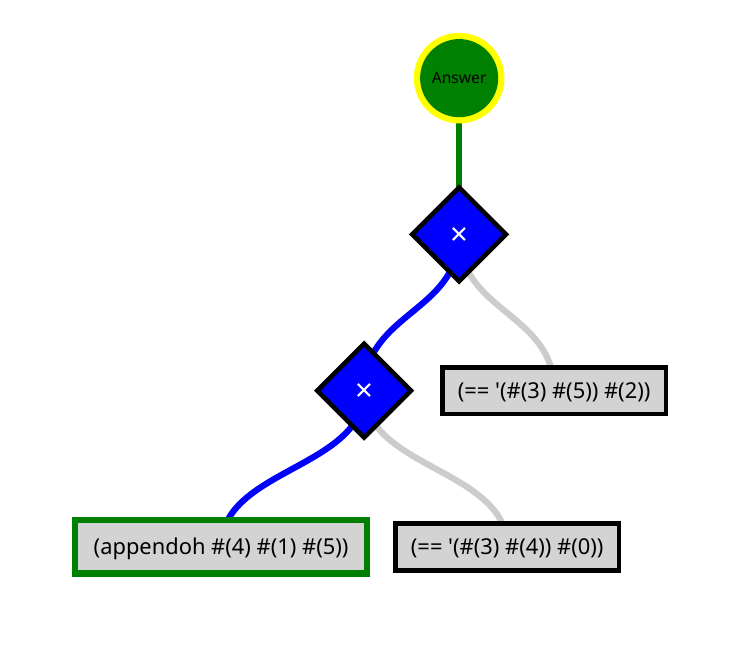}
\end{center}

\noindent While this recursive call did yield another answer, Newb Let notices the goals within the conjunction are still suspended.
\begin{center}
  \includegraphics[width=5cm]{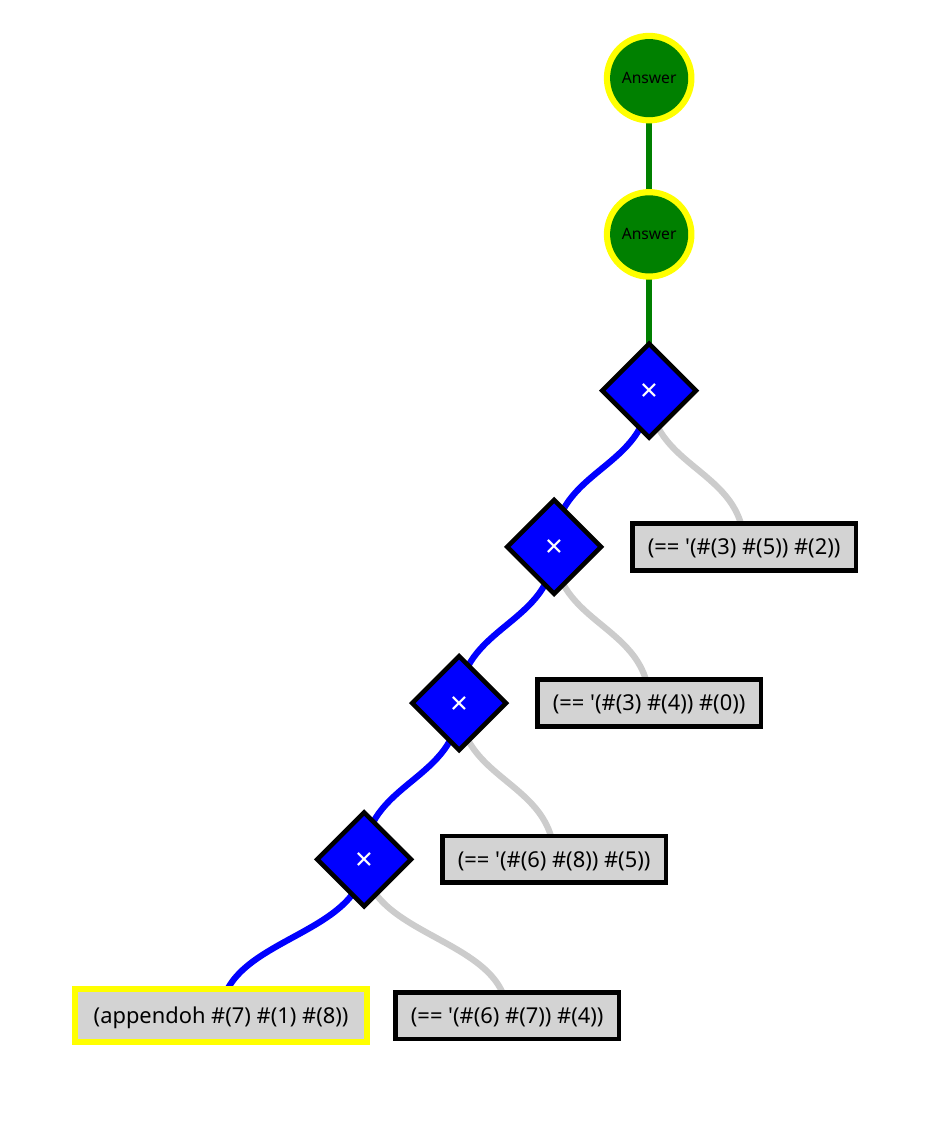}
\end{center}

\noindent As Newb Let continues stepping they get more answers, but it is conjunctions all the way down!
Newb Let realizes that each recursive call results in two more conjunctions which a state has to pass through in order to become an answer.
Interleaving is of no help here, since conjunctions do not permit interleaving.
These conjunctions cannot be pruned either as the recursive call to \texttt{appendoh} continually generates more work to be done.
Newb Let learns a lesson and now places his recursive calls at the bottom of his conjunctions.

\section{Related Work}\label{sec:related-work}

We restrict our focus here to fine-grained visualizations of a logic program's execution---those that attempt to model or expose the dynamic computation process as a search tree, rather than simply annotate a linear trace.
Most early systems did not support real-time visualization, and could only visualize pre-executed computations.
\Citet[\S~2]{brna1991overview} and \citet[\S~7.1]{ducasse1994logic} survey the landscape of such early tools.
In many cases, such a tool is designed as part of a more featureful debugger.
We address here specifically only a few of the most relevant visualizers.

Among the earliest graphical systems in this space is the Transparent Prolog Machine (TPM)~\cite{eisenstadt1988transparent,eisenstadt1990finegrained}, which introduced AORTA diagrams (AND/OR trees~\cite{kowalski1979logic} augmented with status annotations) to depict both the structure of the Prolog search and the control state at each node.
TPM's model combined a static representation of the program structure with a depiction of the current state of the ongoing search process.
That general approach, depicting computation as movement through an AND/OR-like tree, has informed many later visualizers.
TPM itself  was pedagogically oriented and aimed to clarify backtracking and cut behavior.
Later variants implemented a \enquote{live mode} that allowed a user to trace their computation as it ran.
In their approach, each individual image in the visualization was partly-dynamic depiction of computation process over time.
These images displayed both a search tree, and the sequential process through which it was produced.
They indicate that a goal previously failed and was being tried again upon backtracking with so called \enquote{ghost nodes}---depicted as shadows behind the image of the goal in the tree.
Their model also followed Prolog's standard depth-first search and did not permit modification of the search.
\Citeauthor{tamir1995visual}'s PROVIS~\cite{tamir1995visual} targets pure Prolog.
Their system provides a split display of program source and a tree.
In their case the tree is an SLD resolution tree, representing the computation thus far.
Buttons allowed users to interact directly with the GUI.\@
Their visualizer decorates arrows in the SLD tree with the substitution at that point in the computation.
Different substitutions can decorate the arrows along a single path from the root, reflecting the accumulation of knowledge as conjuncts are executed in order.
For performance reasons, they omit the occurs check.
Their tool allowed users to pause execution at disjunctions and explicitly select which branch to explore (so called \enquote{debugging on the tree}), giving users limited control over the search path and enabled them to avoid known-diverging branches.
\Citet{cameron2003vimer} introduce a layered AND/OR representation that captures call structure and backtracking in a dynamic, stepwise fashion.
They represent the current state of the search using an AND/OR style tree, representing prior failed attempts in the third dimension.
A numbered label in the tree indicates the number of previous backtracked states and clicking it makes the prior tree available.
Their tool, ViMer, is a visual debugger for Mercury, incrementally builds a visual representation of execution as the program runs.

Unlike the previous approaches, and ours, \citet{rajan1990principles} and \citet{adachiti1999logichart} do not emphasize the search tree per se.
\Citeauthor{rajan1990principles} describes a single stepper for Prolog called APT that, during dynamic execution, correlates the goal being executed with the relevant lines in the source program.
APT has a status bar to tell the user what operation the trace is currently performing.
Like ours, their visualization is derived from the user's source code, beginning with the query and evolving with the search process.
In the trace representation of the executing query, they use direct substitution both at a clause level (to illustrate a head match) and at the level of individual variables (showing assignments that take place during unification).
They therefore single-step at a sub-unification level.
\Citeauthor{adachiti1999logichart}'s Logichart lays out Prolog clauses and calls in a chart-style diagram aligned with source code structure.
The execution is animated by color-coded highlighting of elements as they are invoked or completed.
This addressed concerns about the abstractness of SLD-trees and aimed to help learners relate behavior to program text.

Our work comes out of a slightly different tradition and with a distinct emphasis.
Rather than treat the search tree as a byproduct of evaluation to be visualized post-hoc or recorded for replay, our evolving search tree is explicitly represented in the program state.
The tree itself is constructed incrementally.
Each tree image corresponds exactly to an intermediate computation state, and every step in the visualizer corresponds 1-1 with a reduction step.
This makes it possible to reflect scheduling and interleaving decisions directly in the visualization, not just as an animated trace, but as state-transforming computation.

\subsection{Operational semantic models for miniKanren}

Several prior efforts have proposed formal or mechanized models of miniKanren execution.
\Citet{rosenblatt2019firstorder} describe a first-order version of miniKanren and used it to implement a stepper that supports user-directed exploration.
Their aim is to produce a tooling-friendly representation of miniKanren's goals and search state.
Their approach supports interactive stepping and human-guided search, but does not formalize the search order or represent the interleaving scheduler explicitly.
The stepping operates at the goal level, as interpreted by an underlying engine, rather than being grounded in a formal reduction semantics.

\Citet{rozplokhas2020certified} present a certified operational semantics for miniKanren, formalized and proved correct in Coq.
Their semantics ensures fair search by construction, but does so via a maximally interleaving strategy that differs in detail from how typical miniKanren systems behave.
That work is oriented toward proof of correctness and completeness, not toward visualization or close-up inspection of the search process.

\section{Conclusions and Future Work}\label{sec:conclusion}

In this work we provide a small-step semantics that models miniKanren's interleaving strategy driven by a focus on implementation behavior and pedagogical inspection.
In practice, both novice and experienced miniKanren programmers encounter surprising behaviors or performance pathologies due to subtle differences in how goals are written or ordered~\cite{rosenblatt2019firstorder}.
The search tree and interleaving control are made explicit in the semantics.
Each visual step reflects a concrete transition in that semantics, exposing interleaving decisions and waiting goals.
We treat the semantics itself as a pedagogical artifact, a notional machine that can be visualized and stepped through.
The resulting tool supports tracing interleaving behavior and reasoning about surprising answer orders, such as those shown in \cref{sec:introduction}.
We imagine continuing this work in several possible directions.

One possible focus of future work would extending this tool into being a full-fledged debugger.
Some of these involve adding new UI features.
Being able to ``fast forward'' or ``back up'' to the next relation call would let a user approximate \citeauthor{rosenblatt2019firstorder}'s goal-level stepping behavior.
Leaving invisible marks for the relations that were called, so you can track the trace and order procedure calls which led to that state.
Our tool currently shows every small step; this yields comprehensive traces but in lengthy executions could be unwieldy---an open question is how to maintain clarity at scale.
One approach would be to add a UI toggle to see ``just the current goal'', ``just the current goal and it's waiting conjuncts'', or the whole search tree.
We could also imagine adding a profiler based around the search tree execution.

One direction we're definitely interested in is user studies to test our hypotheses about this tool's efficacy for students.
User studies and AB-testing could help inform what default UI changes and what new options would help students.
For instance, do students do better with the railroad model of interleaving or actually flipping the interleaved subtree across the vertical axis.
It may be that the tree flipping could make following the transformation more difficult, but at the same time it might clarify the relationship to traditional depth-first search.
The Racket algebraic stepper shows both the before and after of a transformation on the same screen: would that be preferable to our one-tree-at-a-time view?
Currently we represent logic variables in the UI literally \emph{as} numeric constants.
Choices for naming of logic variables that make sense for the reader to look at but do not confuse with names used by the programmer are a tricky issue.
\Citet[\S 4.1]{ducasse1994logic} describe this as a well-known problem.
Still, we think there's room for UI improvement there.
Another question involves the cost of more user-friendly surface syntax.
A great virtue of miniKanren as an embedded DSL is that programmers can invent their own custom language extensions host-language using host-language extension mechanisms~\cite{ballantyne2024compiled}.
Those layers of syntax sugar could, however, open a wide gulf between concise surface syntax using some nice special forms (e.g. \texttt{matche}~\cite{keep2009pattern}), and the nested binary tree notation we present to users.
%

Another future direction would be to continue modeling the interleave position design space.
The miniKanren community has tried out many different varieties of interleave positions.
This implementation approach nicely parameterizes the search behavior, and it would be nice to have all the common choices side-by-side in a common framework that facilitates experimentation.

Finally, we are also interested to explore other semantic artifacts using this language model.
From our reduction semantics, we should be able to follow the refocusing, transition compression, fusion sequence that \citet{danvy2004refocusing} describe, and generate the microKanren big-step abstract machine.
\Citet{biernacki2003interpreter} transform a semi-compositional \enquote{double-barrel} CPSed interpreter for a small Prolog-like language into a logic engine by defunctionalization.
It will be interesting then to both compare our abstract machine with \citeauthor{biernacki2003interpreter}'s, and also to generate both the corresponding microKanren continuation-passing interpreter and natural semantics~\cite{danvy2005reduction}.
These would in turn help extend \citeauthor{wand2004relating}'s~\cite{wand2004relating} work relating models of backtracking to the context of backtracking with interleaving.

\printbibliography{}

\end{document}